\documentclass[conference]{IEEEtran}

\IEEEoverridecommandlockouts
\newcommand{\comment}[1]{}
\usepackage{cite}
\usepackage{amsmath,amssymb,amsfonts}
\usepackage{algorithmic}
\usepackage{graphicx}
\usepackage{textcomp}
\usepackage{tabularx}
\usepackage{booktabs}
\usepackage{xcolor}
\usepackage[hidelinks]{hyperref}
\usepackage{multirow}
\usepackage{subcaption}
\usepackage{nicefrac}

\newcolumntype{C}{>{\centering\arraybackslash}X}

\def\BibTeX{{\rm B\kern-.05em{\sc i\kern-.025em b}\kern-.08em
    T\kern-.1667em\lower.7ex\hbox{E}\kern-.125emX}}
\begin{document}

\title{Binding Dancers Into Attractors}

\author{\IEEEauthorblockN{Franziska Kaltenberger}
\IEEEauthorblockA{\textit{Neuro-Cognitive Modeling}\\
\textit{University of Tübingen}\\
Sand 14, 72076 Tübingen, Germany\\
franziska.kaltenberger@student.uni-tuebingen.de}
\and
\IEEEauthorblockN{Sebastian Otte}
\IEEEauthorblockA{\textit{Neuro-Cognitive Modeling}\\
\textit{University of Tübingen}\\
Sand 14, 72076 Tübingen, Germany\\
sebastian.otte@uni-tuebingen.de}
\and
\IEEEauthorblockN{Martin V. Butz}
\IEEEauthorblockA{\textit{Neuro-Cognitive Modeling}\\
\textit{University of Tübingen}\\
Sand 14, 72076 Tübingen, Germany\\
martin.butz@uni-tuebingen.de}
}

\maketitle

\newcommand{\citep}[1]{\cite{#1}}
\newcommand{\citet}[2]{\cite{#1}}

\begin{abstract}
%
To effectively perceive and process observations in our environment, feature binding and perspective taking are crucial cognitive abilities.
Feature binding combines observed features into one entity, called a Gestalt.
Perspective taking transfers the percept into a canonical, observer-centered frame of reference. 
Here we propose a recurrent neural network model that solves both challenges.
We first train an LSTM to predict 3D motion dynamics from a canonical perspective. 
We then present similar motion dynamics with novel viewpoints and feature arrangements. 
Retrospective inference 
enables the deduction of the canonical perspective.
Combined with a robust mutual-exclusive softmax selection scheme, random feature arrangements are reordered and precisely bound into known Gestalt percepts. 
To corroborate evidence for the architecture's cognitive validity, we examine its behavior on the silhouette illusion, which elicits two competitive Gestalt interpretations of a rotating dancer.
Our system flexibly binds the information of the rotating figure into the alternative attractors resolving the illusion's ambiguity and imagining the respective depth interpretation and the corresponding direction of rotation.
We finally discuss the potential universality of the proposed mechanisms. 

\end{abstract}

\begin{IEEEkeywords}
binding problem, perspective taking, retrospective inference, bistable stimuli, silhouette illusion 
\end{IEEEkeywords}

\section{Introduction}
We are constantly challenged to combine observed stimuli into holistic entities, that is, Gestalten \citep{Koffka:2013}.
How the brain actually accomplishes this algorithmically is yet to be determined \citep{Jaekel:2016}. 
The binding problem \citep{mindBeing_Butz,TheBindingProblem_Treisman} asks how individual features can be bound into a coherent percept, while the perspective may need to be adapted in the mean time \citep{Pavlova:2000}.
In vision, edge contours may need to be bound into coherent shapes, or shapes bound with their respective colors and textures \citep{multiAccBinVis_Humphreys}.
Treisman \cite{TheBindingProblem_Treisman} considered location binding as "the most basic binding problem: linking 'what' to 'where'" (p. 171).
Along related lines, point-light motion displays have been studied intensively over the past decades \citep{Johansson:1973,Pavlova:2012}. In this case the dynamics allow the inference of the cause of the point-light stimuli (e.g. human, dog, bicycle), including, in the case of a human figure, even the weight, (somewhat stereotypic) gender, emotionality, and intentionality \citep{Troje:2002,Pavlova:2000,Pavlova:2012}.

Binding and perspective taking also occurs cross-modally and even cross-conceptually. 
For example, one may bind entities and their dynamic interactions into the concept of preparing tea with milk \citep{Butz:2021,Kuperberg:2021}. 
Binding mechanisms thus play a rather universal role in our brain, being involved on perceptual, conceptual, and even social cognitive levels \citep{mindBeing_Butz}. 
Accordingly, various research strands have developed an integrative,  event-oriented processing perspective, which emphasizes that cross-modal stimulus information and involved entities are bound into Gestalt-like attractors \citep{Zacks:2007,Butz:2016,Butz:2021,Hommel:2001}.
In addition, binding, and thus Gestalt perception, can differ for identical input: 
In the silhouette illusion a rotating two dimensional fully blackended body silhouette is shown, which can be perceived rotating either clockwise or counter-clockwise \citep{Liu:2012}\footnote{Nobuyuki Kayahara designed this illusion: \url{http://www.procreo.jp/labo/labo13}}.
The inferred direction depends on the inferred three dimensional, possibly mirrored, body posture.

Over recent years, focusing on biological motion stimuli, a group of algorithmic models of perspective taking and Gestalt perception were developed \citep{Schrodt:2018,Sadeghi:2021,Sadeghi:2021a}.
These models include the division of visual input into positional, directional, and motion magnitude information as well as their respective encoding via population codes. 
The population-encoded stimuli are compressed into latent Gestalten by means of an autoencoder, which projects the Gestalten back onto the sensory input stream.
After sufficient learning of the considered Gestalten, 
retrospective inference \citep{Butz:2019,ActiveTuning_Otte2020} allows the deduction of perspective and Gestalt-oriented bindings.

In contrast to these closely related approaches \citep{Schrodt:2018,Sadeghi:2021,Sadeghi:2021a}, our model fully focuses on encoding temporal dynamics and does not consider population encodings. 
In particular, we introduce a cognitive system that solves the perspective taking and binding challenge when facing dynamic motion stimuli. 
Moreover, the system models the silhouette illusion, thus underlining its cognitive plausibility. 
While relying on standard backpropagation through time when learning temporally predictive Gestalt encodings, the system solves the challenges by means of retrospective inference \citep{Butz:2019}, which is inspired by theories of predictive coding and free energy minimization \cite{PredCod_Friston}.
In our perceptual task, free energy minimization essentially adapts its recurrent dynamics concurrently with internal parametric biases \cite{Sugita:2011a}, which control feature binding and the taken perspective, for minimizing prediction error. 
In the case of the silhouette illusion, we show that the system can infer either one of two dancer interpretation attractors and can switch between them.
Our approach may be applied to other binding challenges, including scene interpretations, cause-and-effect conceptualizations, or role-assignments in language.


\section{Adaptive Gestalt perception model}

\begin{figure*}
\centering
\includegraphics[]{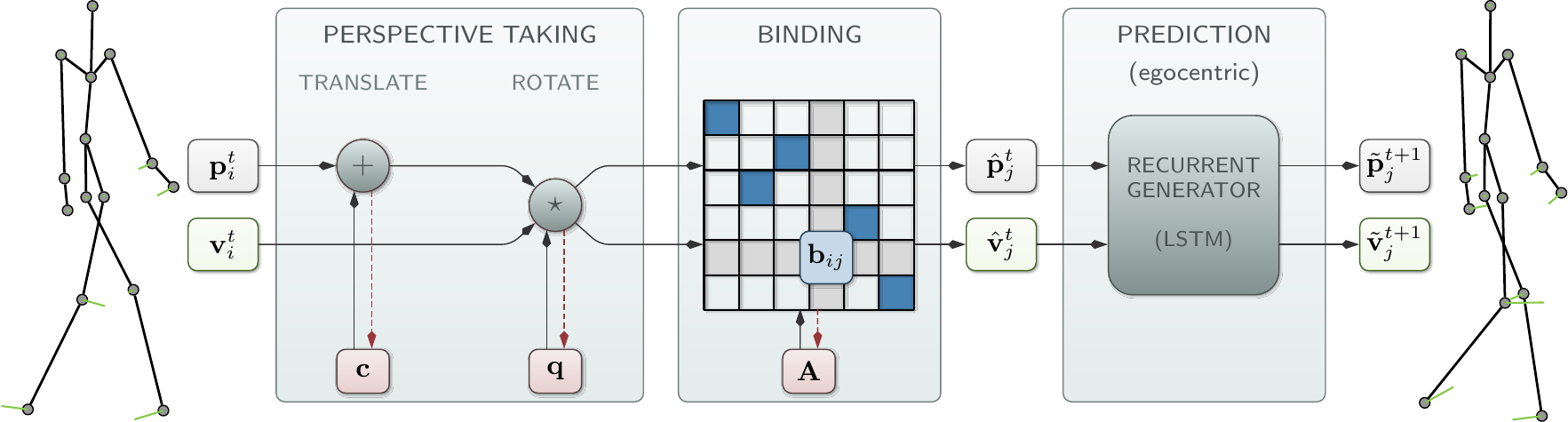}
\caption{The adaptive Gestalt perception architecture: Position vectors $\mathbf{p}_{i}^{t}$ and velocity vectors $\mathbf{v}_{i}^{t}$ of a set of posture markers serve as input. The vectors first undergo a perspective taking stage, 
involving a translation of the reference frame by a vector $\mathbf{c}$ and a feature-oriented, quaternion-based rotation $\mathbf{q}$. The transformed features are then bound onto respective inputs of a recurrent motion predictor, which was trained on egocentric, canonical motion dynamics. The involved binding matrix is parameterized with activities $\mathbf{A}$, which are passed through row-wise and column-wise softmax operations to foster mutually-exclusive feature binding values $\mathbf{b}_{ij}$. Red dotted lines indicate gradient signals used to adapt perspective and binding.
}

\label{fig:model}
\end{figure*}

In this section we present our adaptive Gestalt perception model and its  components. The overall architecture is depicted in Figure~\ref{fig:model}.
As input we provide set of position and velocity vectors, $\mathbf{p}$ and $\mathbf{v}$, of multiple skeleton markers. In the following we refer to one such position and velocity pair as a feature.

The core of our model is an LSTM-based \cite{hochreiter1997long} motion predictor that learns the contextual assignment and interpretation of the body input features from an egocentric, canonical perspective and motion sequences of the own body, where all body parts are known and ideally assigned. Using this embodied prediction model to capture and predict Gestalt patterns and their dynamics of another entity involves two essential preprocessing steps, namely, perspective taking and feature binding. 

Perspective taking enables humans to recognize Gestalt patterns in the motion of an observed person by mentally adopting the spatial perspective \citep{EmbodiedSPT_Kessler}. Spatial perspective taking could be interpreted as a "mental rotation of the self" (p. 73) from an egocentric reference frame into the perspective of an observed person. 

Feature binding enables the perception of observed features as one entity, allowing a temporal and contextual interpretation of the observed input \citep{TheBindingProblem_Treisman}. Feature binding here addresses the task of optimally routing information of observed non-assigned (unsorted) features to the corresponding, that is, best matching input neurons of the core LSTM network, which corresponds to the most plausible Gestalt perception.

\subsection{Perspective taking}

Perspective taking requires two transformations, namely, translation and rotation. The translation of the posture Gestalt is performed by adding the current translation vector $\mathbf{c} \in \mathbb{R}^3$ to the current positional vector $\mathbf{p}_{j}^{t}$ for all observations $j$. We implement the rotation using a unity quaternion $\mathbf{q} = (w, x, y, z)$ with $\vert\mathbf{q}\vert=1$  (cf. e.g. \citep{jia2008quaternions}). We denote the rotation of a vector $\mathbf{u}$ by the quaternion  $\mathbf{q}$ with $\mathbf{q} \star \mathbf{u}$. Quaternions provide several advantages when dealing with rotations in the context of neural networks \cite{traub2021manyjoint}. Preliminary experiments indicated that quaternions work significantly better than Euler angles, which were employed previously \cite{Sadeghi:2021}. 
The rotation operation is performed for all velocity observations $\mathbf{v}_{j}^{t}$ as well as all translated positional observations $\mathbf{(\mathbf{p}_{j}^{t} + \mathbf{c})}$.
Both $\mathbf{c}$ and $\mathbf{q}$ are adapted by gradient signals using retrospective inference. Note that we force unity of $\mathbf{q}$ immediately after each adaption step.

\subsection{Feature binding}

We realize feature binding using a binding matrix $\mathbf{B} \in [0, 1]^{N\times N}$, where $N$ is the number of observed features (which we for now assume to be equal to the number of canonical features, here 15). The binding matrix is multiplied with the observational input, proportionally binding the observations to the input features of the core LSTM network. Each value $b_{ij}$ expresses how strong the observed feature $j$ will be associated with the canonical feature $i$. An ideal binding matrix provides sharp, unique, and exclusive assignments. 

$\mathbf{B}$ is calculated using a sequences of operations starting with a matrix of raw binding activities $\mathbf{A} \in \mathbb{R}^{N \times N}$, which we infer using retrospective inference. In contrast to previous approaches in which $\mathbf{B}$ was obtained by applying the standard sigmoid function component-wise to the raw binding matrix (to ensure $0 \leq b_{ij} \leq 1$), we propose a softmax-based solution. Recall that the softmax for the vector $\mathbf{x}\in \mathbb{R}^{n}$ is calculated with:%
\begin{equation}\label{eq:rwSM}
    \operatorname{softmax}(\mathbf{x}) = \left(
        \frac{
            \exp\left(
                \displaystyle
                \frac{x_{i}}{\tau}
            \right)
        }{
            \sum_{i'} \exp\left(
                \displaystyle
                \frac{x_{i'}}{\tau}
            \right)
        }
    \right)_{i = 1, \ldots, n},
\end{equation}\label{eq:softmax}%
where the temperature variable $\tau$ regulates how smooth or sharp the resulting vector is and, thus, how the softmax contrasts numerical differences within the input \cite{hinton2015dist}. Low temperature values $0 < \tau < 1$ amplify numerical differences between the input values, whereas high temperature values suppress them.

To foster the exclusive binding of individual stimuli, we first apply the softmax row-wise on $\mathbf{A}$, yielding matrix $\mathbf{B}^{rw}$. This stresses the selection of the best matching observation feature $j$ for each canonical feature $i$ (feature selection). 
Second, we apply the softmax column-wise on $\mathbf{A}$, resulting in the matrix $\mathbf{B}^{cw}$, enhancing the selection of the best matching canonical feature $i$ for every observation $j$ 
(feature exclusion). This fosters a mutually exclusive feature assignment: each observation feature $j$ will be assignable to only one canonical feature $i$.

The final binding values in $\mathbf{B}$ are a result of combining $\mathbf{B}^{rw}$ and $\mathbf{B}^{cw}$ via the square root values of the Hadamard product (element-wise multiplication):
\begin{equation}\label{eq:optimal_combination}
    b_{ij} = \sqrt{b_{ij}^{cw} \cdot b_{ij}^{rw}},
\end{equation}
where the square-root compensates the ``square effect'' caused by the Hadamard product. 
We use individual temperature values $\tau^{rm}$ and $\tau^{cw}$ for the row-wise and column-wise softmax, respectively, controlling their respective exclusive binding strengths.
A large temperature results in an almost uniform binding matrix, i.e. all observation features are assigned to all canonical features. Decreasing the temperature decreases the binding uncertainty, moving towards a sharp, exclusive assignment.
This binding mechanism thus allows focusing attention, which has been considered essential for solving the binding problem also from a neuroscience perspective \citep{friedenberg2011cognitive,reynolds1999role,FeatInTheo_Treisman, treisman1998feature}.

\subsection{Asymmetrical N$\times$M binding}

In case of the previously described symmetrical $N \times N$ binding, all observed features are assumed to correspond to one canonical input feature. When there are more observations than canonical features, we need an asymmetrical $N\times M$ binding with $M > N$: $M$ observed motion features are bound onto $N$ input features of the core LSTM network, which inevitably requires the exclusion of some (irrelevant) observation features. In this case, however, the column-wise softmax may amplify irrelevant features possibly then even winning the tendency toward mutual exclusion.
To alleviate this problem, we extend the raw binding matrix with an additional row, which allows marking features as irrelevant, excluding them from further processing.
Consequently, we refer to this line as an outcast line. 
The outcast line does not undergo the row-wise softmax computation to enable the exclusion of several features from further stimulus processing.

\subsection{Adaptive perception through retrospective inference}
%


In order to accomplish flexible, adaptive binding and perspective taking functionality in our architecture, that is, inferring $\mathbf{A}$, $\mathbf{c}$, and $\mathbf{q}$, we apply retrospective inference \cite{butz_learning_2019,ActiveTuning_Otte2020}. 
This gradient-based inference process tunes parametric biases \cite{Sugita:2011a} and other latent neural states to minimize prediction error over time.
To do so, it back-propagates prediction error-induced temporal gradient signals over a retrospective temporal horizon of length $H$. 
Interleaved with forward roll-outs, the latent state adaptions yield progressively more accurate predictions.

In particular, we used smooth L1-Loss, which is projected into the past along the tuning horizon by means of backpropagation through time \citep{werbos1990backpropagation}. 
As a result, state variables $\mathbf{A}$, $\mathbf{c}$, and $\mathbf{q}$ are adapted by gradient descent (with respective learning rate $\eta$ and momentum rate $\mu$). The resulting parameter adaptation signals are averaged over the tuning horizon.
In addition, adaptation signals for perspective taking ($\nicefrac{\partial \mathcal{L}}{\partial\mathbf{c}}$ and $\nicefrac{\partial \mathcal{L}}{\partial\mathbf{q}}$) are scaled by $\nicefrac{1}{N}$.
Based on the adapted binding and perspective talking variables, the model is then rolled out again, resulting in a corrected sequence of predictions. 
One iteration of the described procedure is referred to as a tuning cycle realizing one inference step. For every reference time step, multiple tuning cycles can be performed. 

During initial experiments we discovered strong fluctuations of the gradient signals, which sometimes affected the convergence of the inference procedure. To alleviate this problem, we applied a technique called sign damping \cite{otte2018integrative}. 
For each component of the gradient, here written as $g_{i}$, a low-pass filter on its sign across the inference steps is applied:
\begin{equation}\label{sign_damp}
    s_{i}^k = \alpha s_{i}^{k-1} + (1-\alpha) \operatorname{sgn}(g_{i}^{k})
\end{equation}
where $k$ defers to the current inference step and $\alpha \in [0,1]$ determines how strong the sign is smoothed over time. Before it is used for optimization purposes, the gradient $g_{i}$ is scaled by $(s_{i}^{k})^{2}$. In case of a strongly oscillating gradient $(s_{i}^{k})^{2}$ becomes very small and thus effectively tunes down the amplitude of the gradient. Otherwise $(s_{i}^{k})^{2}$ remains close to 1.

\section{Experiments}

We evaluate our model on biological motion dynamics of a walking human as well as on the silhouette illusion. 
In particular, to quantify the general ability to bind perceptual stimuli into Gestalten, and to adjust the perspective onto a Gestalt where necessary, we tackle binding and perspective taking tasks on real motion capture data, similar to \cite{Sadeghi:2021a}.
Moreover, we model the bistable perception of the silhouette illusion.
In this case, during inference, we present a two-dimensional stimulus while the model infers the bistable depth information, effectively imagining the three dimensional figure from the two dimensional stimulus information.


To evaluate model performance quantitatively, we 
compare the values of the inferred binding matrix $\mathbf{B}^k$ with the optimal matrix $\mathbf{B}^{opt}$: 
\begin{equation}
    FBE(k) = \sum_{j=1}^M \sqrt{\sum_{i=0}^N \left(b_{ij}^k - b^{opt}_{ij}\right)^2 },
\end{equation}
effectively computing a \emph{feature binding error}. 
To assess perspective taking performance, we compute a \emph{rotation error} $RE$ as the angular difference (in degrees) between an ideal rotation quaternion $\mathbf{q}^{opt}$ and the updated quaternion $\mathbf{q}^k$ 
by:
\begin{equation} \label{eq:roterr}
    RE(k) = 2 \arccos \left( \vert \langle \mathbf{q}^{opt}, \mathbf{q}^k \rangle \vert \right)
\end{equation}
Finally, the \emph{translation error} between the ideal translation vector $\mathbf{c}^{opt}$ and the current translation vector $\mathbf{c}^k$ is defined as:
\begin{equation} \label{eq:transerr}
    TE(k) = \Vert \mathbf{c}^{opt} - \mathbf{c}^k \Vert.
\end{equation}
The settings for the model's hyper-parameters are given in Table \ref{tab:exp_hyperpars}. 
While most parameters were not particularly critical, it proved to be important to choose a sufficiently long tuning horizon $H\geq 10$, which indicates that indeed longer-term temporal dynamics are encoded and crucial for an accurate inference of Gestalt percepts.


\begin{table*}[t!]
\caption{Model hyperparameters for performed experiments.}
\begin{center}
\begin{tabularx}{\linewidth}{ rCCCCCCCCCCCCCC} 
\toprule
\multirow{2}{*}{\vspace{-0.13cm}\textbf{Experiment}} &
\multicolumn{14}{c}{\textbf{Model hyperparameters}} \\ 
\cmidrule{2-15}
 & $H$ & $cyc$ & $\sigma$ & $\tau^{rw}_{min}$ & $\tau^{cw}_{min}$ & $\alpha^b$ & $\alpha^r$ & $\alpha^t$ & $\eta^b$ & $\eta^r$ & $\eta^t$ & $\gamma^b$ & $\gamma^r$ & $\gamma^t$  \\ 
\midrule

\textit{Walker (symmetrical)} & 10 & 1 & 7 & $1/3$ & $1/5$ & 0.7 & 0.9 & 0.9 & 1.0 & 0.01 & 0.05 & 0.9 & 0.4 & 0.4  \\ 
\textit{Walker (asymmetrical)} & 10 & 1 & 7 & $1/3$ & $1/5$ & 0.7 & 0.9 & 0.9 & 1.0 & 0.01 & 0.05 & 0.95 & 0.4 & 0.4 \\ 
\textit{Silhouette illusion} & 10 & 4 & 5 & $1/3$ & $1/5$ & 0.5 & - & - & 1.0 & - & - & 0.95 & - & -  \\
\bottomrule
\end{tabularx}
\label{tab:exp_hyperpars}
\end{center}
\end{table*}

\subsection{Binding and perspective taking for walking motion}
The walker experiments are based on motion capture data from the Carnegie Mellon University \citep{cmu_data}.
30 three dimensional body features were recorded with a frequency of 120Hz and a resolution of 4 megapixel using 12 infrared MX-40 cameras and 41 body markers. 
We used 15 of the 30 features similar to \cite{Sadeghi:2021a}.
The core LSTM network was trained only on one particular motion sequence of 900 frames sampled at 120Hz (subject 35, trial 7), which we refer to as \emph{known data}. 
During inference, two additional motion sequences (subject 5, trial 1 and subject 6, trial 1), which we call \emph{unknown data}, are used. 
The recorded subject is walking in all sequence samples. 
If the number of frames differs between samples for an inference task, shorter motion sequences are repeated, so that the number of frames is identical for all samples.
All positional data obtained from the motion capture sample sequences was scaled into a range of $[-1, 1]$.

During training, the network receives data optimally bound in an egocentric context from the 900 frames of known data.
Random noise uniformly drawn from $[-2\cdot 10^{-5}, 2\cdot 10^{-5})$ was added to all feature values in the input frames, while target frames remained unchanged, thus training the LSTM module as a denoising system. 
Input batches contained 10 consecutive frames.
In every epoch, $\lfloor \frac{2}{3} \rfloor$ of all batches were selected randomly. 
We used a single LSTM Cell with 100 hidden units and a linear output layer.
Training ran for 2\,000 epochs with the MSE loss and the Adam optimizer \citep{kingma2014adam} (all parameters set to default) using a learning rate of $0.01$.
During training, the MSE error decreased from approximately 0.04 to a value of $1.8\cdot10^{-6}$ after 2\,000 epochs. 
When probing closed loop behavior of the system at the end of training we observed stable imaginary walking behavior for over 70 time steps, when initializing the latent state via teacher forcing for 10 frames.

\begin{figure*}[t]
    \centering
    \begin{subfigure}[b]{0.59\textwidth} 
        \centering
        \includegraphics[width=\linewidth]{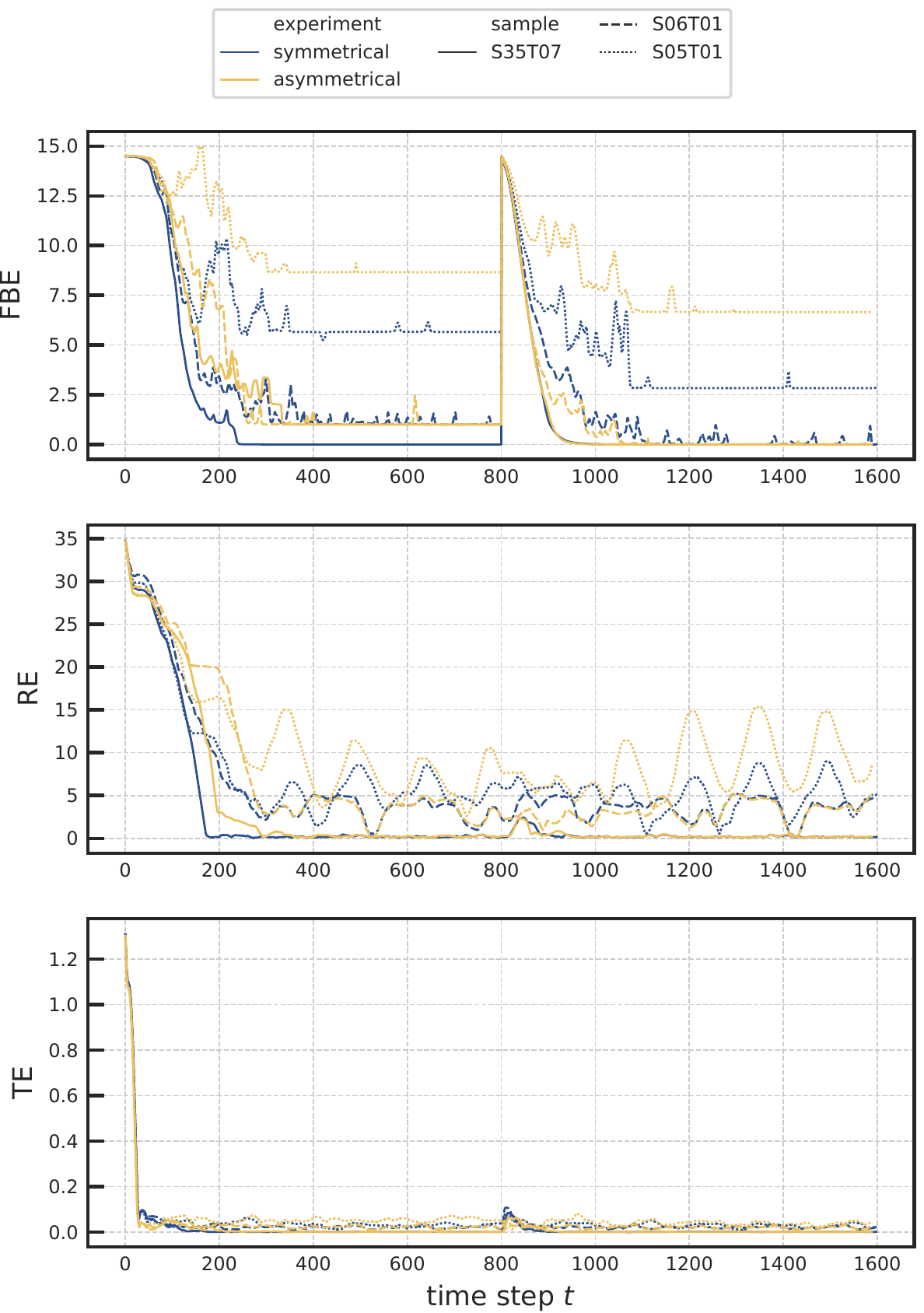}
        \caption{}
        \label{figure:walker_resultsA}
    \end{subfigure}
    \hfill
    \begin{subfigure}[b]{0.39\textwidth}
        \centering
        \includegraphics[width=\linewidth]{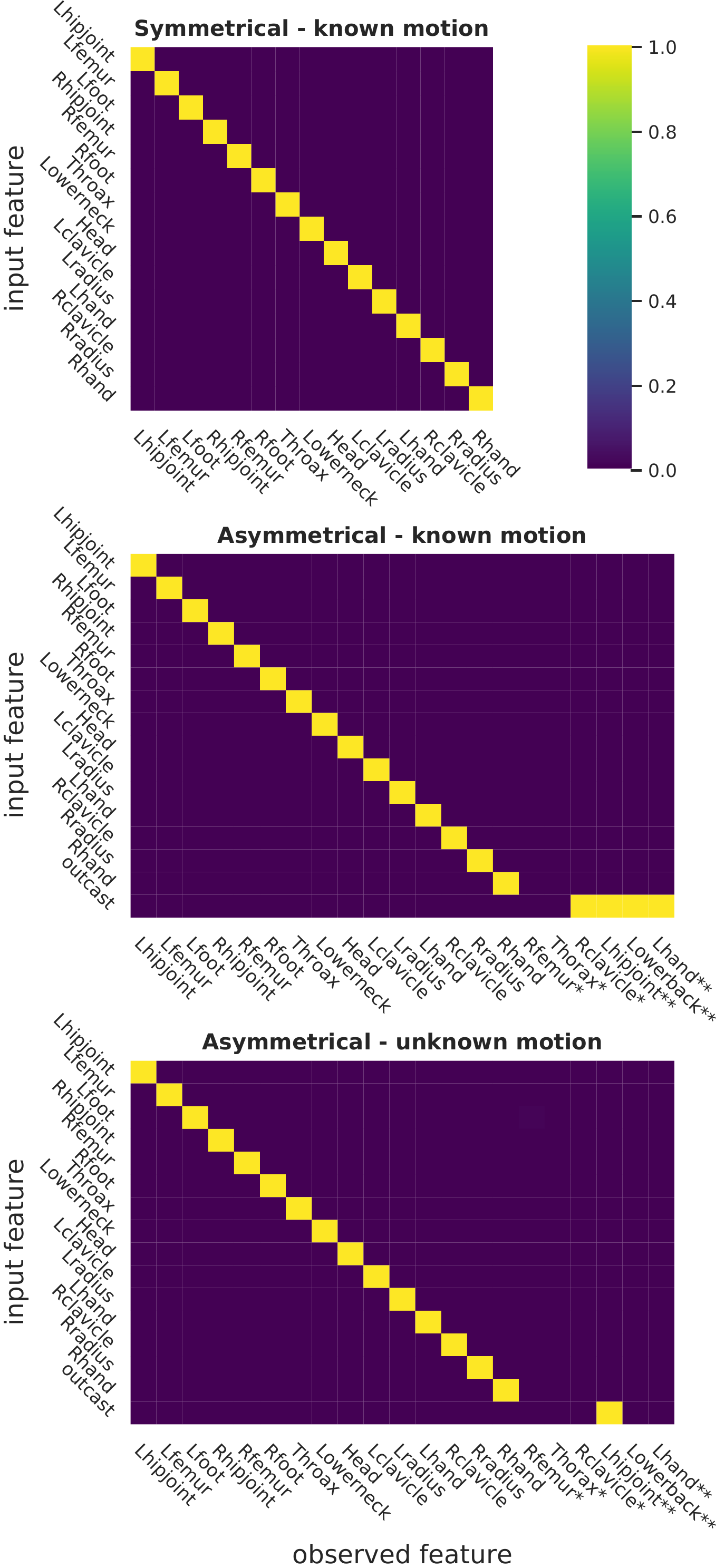}
        \caption{}
        \label{figure:walker_resultsB}
    \end{subfigure}
    \caption{
    Inference results for binding and perspective taking on walking motion. (a): Evolution of feature binding error $FBE$ of the $N \times N$ binding matrix (top), rotation error $RE$ (middle), and translation error $TE$ (bottom) during inference. (b): Final binding matrix $B$ for symmetrical binding of known motion (top), as well as asymmetrical binding of known motion (middle) and unknown motion sequence S06T01 (bottom). Distractors for asymmetrical binding were taken from thai chi (*) and modern dance (**) motion. 
    }
    \label{}
\end{figure*}

Main inference results are shown in Figure~\ref{figure:walker_resultsA}.
The dark blue lines show runs without additional distractor stimuli, while the light orange ones depict inference performance when six additional dynamic distractor stimuli were added. 
As distractor stimuli we took one arm, one torso and one leg feature from a thai chi and a modern dance motion sequence\footnote{Also taken from the CMU database from subject 12, trial 4 (thai chi) and subject 5, trial 2 (modern dance).}. 
Dotted lines depict inference performance on the two unknown data sequences. 

To progressively focus the binding mechanism, we initialized the temperature with $\tau^0=1\,000$ and progressively annealed it to $\tau^{rw}_{min}=1/3$ and $\tau^{cw}_{min}=1/5$, for the row-wise and column-wise softmax operations, respectively. 
For every time step $t > 0$, the current temperature $\tau^t$ of the respective softmax operation is calculated by: 
\begin{equation}\label{eq:temperature}
    \tau^t = \frac{\tau_{max}}{1 + \lambda t},
\end{equation}
using $\tau_{max}$ and $\lambda$ to adapt the speed of temperature annealing.
These annealing parameters were set to $\tau_{max}=530, \lambda=4.6$ for symmetrical and $\tau_{max}=550, \lambda=4.7$ for asymmetrical binding.

The model succeeds in correctly binding the observed features---perfectly for the known data sequence and equally well for one unknown data sequence (S06T01). 
For the second unknown motion sequence (S05T01), binding converged on a higher value.
This difference between motion sequences can also be observed in perspective inference.
However, rotation error declines to a small value after approximately 300 time steps and translation error converges close to 0.0 after 200 time steps for all sequences.
Clamping binding activations to the range $[-\sigma, \sigma]$, as well as the speed of temperature annealing proved to be essential to ensure the parallel inference of all parameters.

At inference step 800, the temperature is reset to $\tau^0$ and decreased again to the minimal respective values.
The results indicate that the model can re-infer stable bindings faster than before and even improve them, as translation and rotation nearly remain unchanged.
It appears that the prediction error yielding re-annealing of the binding matrix falls perfectly into the known Gestalt, thus hardly sending any meaningful error signal down to the rotation and translation parameters.

Asymmetrical binding poses a slightly harder challenge. 
Before the temperature is reset, no optimal binding can be achieved. 
For the better two motion sequences, evaluation of the corresponding binding matrices (not shown here), however, shows that the lower back of the walker is not assigned to any input feature of the LSTM model, while remaining observations are indeed optimally bound. 
After temperature reset, all features are optimally bound.

Figure~\ref{figure:walker_resultsB} shows final binding matrices for both symmetrical and asymmetrical binding contexts. 
The $N\times N$ binding matrix looks absolutely perfect yielding the identity matrix, even when the data is unknown.
In the case of the asymmetrical binding context, most of the distractors are assigned to the outcast line with minimal variations. 

Overall, the results imply that the model finds a stable internal attractor state even if the dynamics of the observed data differs from the learned motion (in the case of the unknown walker data). 
We leave it for further research to examine the necessary degree of similarity between known and similar, unknown data, which certainly also depends on the amount of denoising during training, the exact RNN architecture employed, and the diversity in the training data.

\subsection{Modeling the silhouette illusion}

The silhouette illusion shows the rotating two-dimensional silhouette of a female dancer, which is perceived as a three-dimensional figure. 
The figure is perceived to rotate either clockwise or---the mirrored version of her---counterclockwise, effectively yielding a bistable percept. 
The two percepts only differ in their assignment of the differently bent and stretched limbs, mirroring the left and right side of the body and inferring stimulus-matching depth information dependent on the current percept. 

To model the silhouette illusion, positional information for the same body features that were used in the walker experiment was scripted for a basic dancer posture. 
Starting with this posture, four versions of a rotating dancer were constructed. 
The basic dancer posture rotates around the height-axis by approximately 4 degrees between every frame, yielding 90 frames for a full rotation.  
This rotation was conducted counter-clockwise and clockwise resulting in two dancer versions ($D+$ and $D-$).
For the other two versions, the basic dancer posture was mirrored in arm and leg features. 
This was realized by negating the positional width information of the respective features and assigning it to the opposite body side. 
The positional information of the torso remained unchanged. 
The mirrored dancer posture was also rotated counter-clockwise and clockwise, resulting in the third and fourth ($E+$ and $E-$) dancer version.


The employed LSTM network consisted of a single LSTM Cell with 300 hidden units and one linear output layer.
Training was performed on the 4 times 90 frames for the four dancer version. 
An epoch comprised 9\,000 frames for each dancer. 
Training ran for 500 epochs using a learning rate of $0.01$ and adding uniform random noise in $[-1\cdot 10^{-4}, 1\cdot 10^{-4})$ to the input but not to the target output values.
In every epoch the network was presented with individual batches, which contained 20 consecutive frames, one for each dancer version. 
After training and teacher-forcing-based initialization for $20$ frames, the behavior of all four dancers is reliably simulated for at least one full rotation, that is, $90$ frames. 

The inference evaluation aims at investigating the adaption behavior to different dancer-respective attractors.
Throughout the experiment we only provide $x$ and $y$ input and infer the depth information $z$---equivalently to the silhouette illusion setup. 
The temperature values are initialized with $\tau^0=1\,000$ at the start of the experiment and progressively annealed to respective minima $\tau_{min}^{rw} = 1/3$ and $\tau_{min}^{cw} = 1/5$.
Temperature annealing parameters are set to $\tau_{max}=130$ and  $\lambda=1.0$.
First, we fix the raw binding matrix parameters $\mathbf{A}$ to the diagonal and observe the predicted $z$ values as well as the general convergence towards the respective attractor.
After $80$ frames we then release the binding matrix parameters $\mathbf{A}$, adapting them via retrospective inference from then onward. 
After $200$ frames, we then feed in the matching depth information $z$ for the opposite dancer version, but only for the left hand input feature and only for $80$ frames. 
Simultaneously, the temperature values are reset to the respective initialization value $\tau^0$ and decreased as specified above.
We continue running the system in depth imagination mode for the rest of the experiment.  

\begin{figure*}[htbp]
    \centering
    \hspace*{-0.8cm}\includegraphics[scale=0.99]{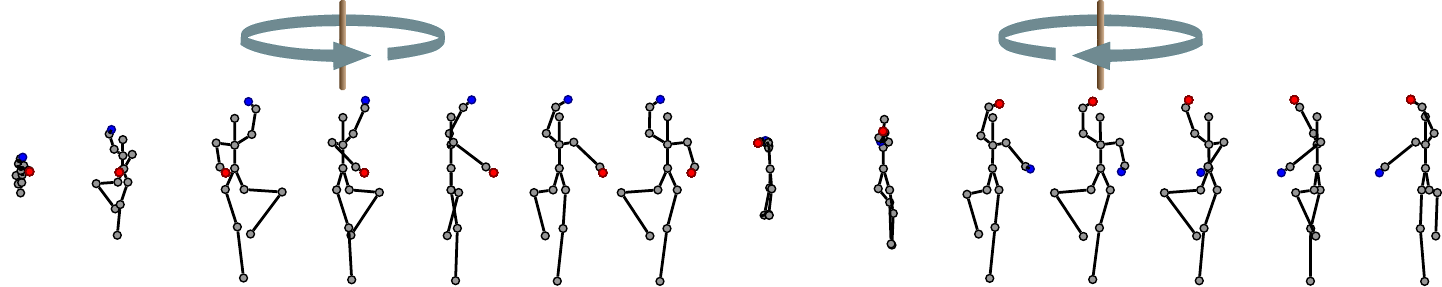}\\
    \vspace{0.2cm}
    \includegraphics[width=\textwidth]{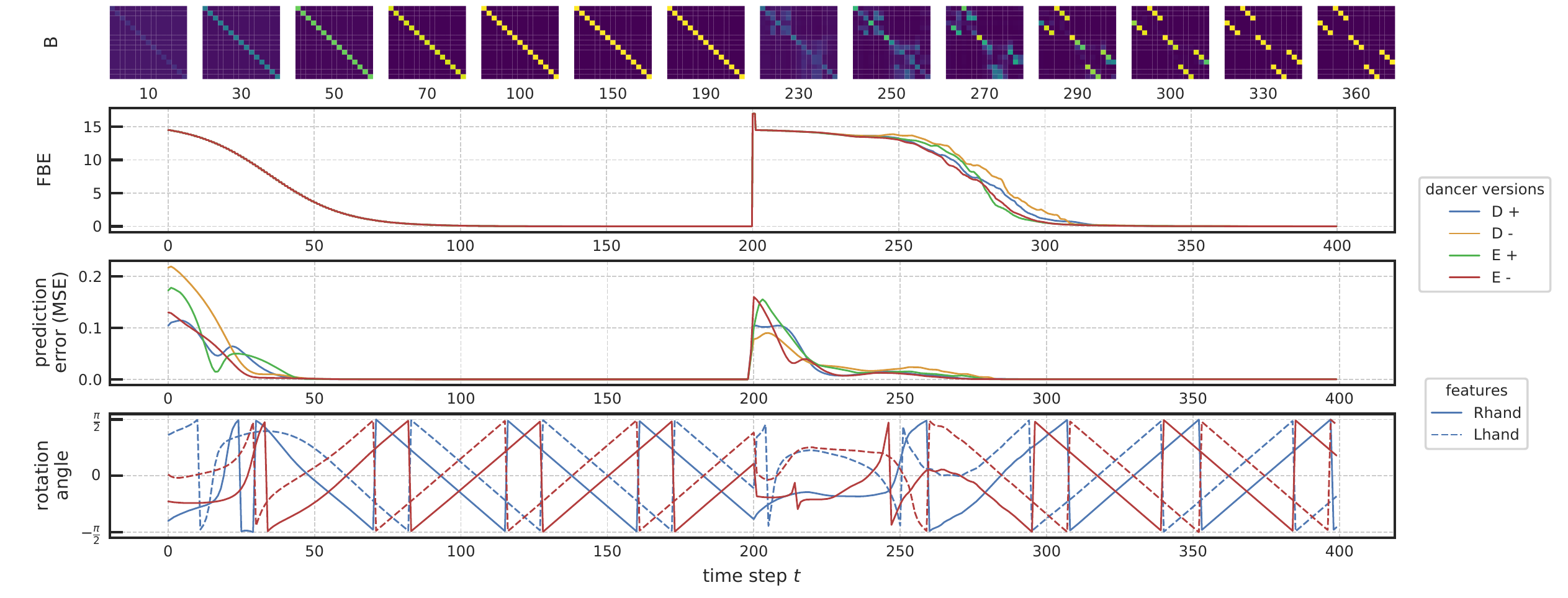}
    \caption{Modeling the silhouette illusion. 
    Development of dancer postures and binding matrix $\mathbf{B}$ over time is shown for initial dancer version $D+$ holding up the left hand (blue) and the right hand (red) in front, then switching to dancer version $E-$.
    For improved illustration purposes, dancer skeletons and binding matrices are not fully aligned with the error and prediction plots. 
    For the first 80 frames, binding activations in $\mathbf{A}$ are set according to one dancer version. From then onward, $\mathbf{A}$ is adapted via retrospective inference.
    After 200 frames a depth cue only in the left hand implies the alternative dancer interpretation. 
    Optimal feature assignments are perfectly maintained for the first 200 frames. Thereafter, the z-axis signal combined with the increased temperature leads to the emergence of the alternative, mirrored mapping. 
    The feature binding error pattern (top) confirms this interpretation. 
    Prediction errors (MSE) for predicted Gestalt values (middle) and corresponding rotation angles in the horizontal plane around the vertical rotation axis (bottom) converge before $FBE$, indicating that the model first switches internal posture and rotation direction before fully optimizing the binding activations.}
    \label{figure:dancer_results}
\end{figure*}

Figure~\ref{figure:dancer_results} (upper two plots) shows the development of the binding matrix $\mathbf{B}$ for one of the dancer types and the $FBE$ for all four dancer types over time. During the first $80$ frames, the temperature annealing progressively reveals the set diagonal activations. From then onward, though, it can be observed that the inference does not negatively affect the binding matrix but the feature binding error continues to converge toward zero. 

Notably, the closed-loop prediction behavior for the four dancers only slightly differs. 
After 200 frames, the induced depth information in the left hand of the opposite dancer causes the expected confusion in the binding. 
At first, binding inference tends to fall back into the previous binding assignment (i.e. complete diagonal binding) since activations themselves are not reset, only the temperature is strongly increased. 
However, already after about 50 time steps, the $FBE$ declines quickly back to zero and yields the mirrored binding (switching left and right arms and legs), which corresponds to the perfectly mirrored dancer posture.

The $FBE$ results, however, do not show yet, whether the internal representation also switches motion rotation direction. The switch direction of rotation is shown in the depth prediction angle plot (Figure~\ref{figure:dancer_results}). Here, the rotation angles of the right and left hand are shown with respect to the horizontal plane. Initially, the rotation angles indicate the search for the correct attractor. However, already after only about 30 frames the angles fall into the correct interpretation: 
dancer $D+$ rotates clockwise while the mirrored dancer $E-$ rotates counterclockwise. 
At frame $200$, the rotations stall, seeking for their respective novel roles. 
After approximately $50$ time steps the opposite directions are picked-up, confirming that the system indeed imagines the opposite rotation direction of the whole dancer. 
This confirms that the architecture can reproduce the bistable perceptive interpretation involved in the  silhouette illusion. 

Finally, the prediction error dynamics further confirm these observations. 
Interestingly, the error improves much earlier than the binding matrix adaptation and the $FBE$, indicating that the model first switches internal Gestalt-like representations and then explains-away the remaining residual error by optimizing the binding activations to the mirrored version. 

Overall, the presented experiment shows that the interpretation of the silhouette can be actively switched by focusing on a specific feature assignment via the binding matrix or via a feature-specific depth cue. 
Very similar phenomena can be noticed when observing the actual silhouette illusion, indicating that our model indeed closely mimics the computational processes and inference dynamics unfolding in our brains when perceiving the ambiguous  stimuli eliciting the silhouette illusion.

\section{Summary and conclusions}

As demonstrated by the presented experiments, our cognitive architecture succeeds in binding observed body features of point-like motion into different attractor conditions.
It manages to infer optimal bindings while simultaneously adapting to the perspective of the observations when these differ from the canonical frame of reference the system was trained on. 
Our model is able to reach a stable attractor state by inferring suitable parameters for binding and perspective taking for known and unknown motion sequences.
Furthermore, distractors can be successfully excluded by assigning them to an outcast binding row.

In addition, the model simulated the bistable perception of the silhouette illusion. 
Internal representations can be switched by inducing opposed depth interpretation in one feature only. 
Feature binding is then tuned accordingly into the corresponding perception.

In contrast to previous, neural network-based cognitive models of the binding and perspective taking challenges \citep{Schrodt:2018,Sadeghi:2021,Sadeghi:2021a}, the presented approach solves these challenges by focusing on the temporal regularities only, omitting population codes and autoencoders. 
The encoding of the data into position and velocity information was useful, though, as was the introduction of a new focusing mechanism. 
The softmax combination, aligned with the temperature parameter and the clamping of binding activation values, ensures good, sharp, and stable binding in various contexts, introduces a concept of gradual binding uncertainty, and allows the model to adapt the binding of its perceptual Gestalt to provided attractors.

Besides the ability to model bistable perceptions, the evaluations on the Silhouette illusion show that our model is applicable to multiple scenarios and different kinds of input data.
The omission of the population codes allows the application of the model also to non-spatial data. 
Data types could even vary within the input allowing for cross-modal event bindings.

At the moment our architecture needs to be trained with canonical feature assignments. 
From an embodied cognition perspective, such an assignment may correspond to the perception of the own body. 
However, when considering other stimuli, such as the perception of other animals or objects, it may be necessary to impose some default assignment of individual features to input channels during learning. 
These assignments could directly depend on local proximity of the features and could furthermore incorporate other information sources, such as apparent feature connectivity. 

In the near future, we hope to combine the temporal Gestalt encoding mechanism with the previously employed static autoencoder-based Gestalt encodings, possibly reaping inference benefits from both, static, constellation-oriented and dynamic, spatial-relational information sources.
This should enable the simulation of other bistable illusions. 
More importantly, though, it may pave the way towards an even broader application of the introduced binding mechanism also to other binding challenges, such as binding entities and interaction dynamics into events or binding words to available roles in a grammatical sentence.  

\section{Acknowledgements}
Walking motion data used in this project was obtained from mocap.cs.cmu.edu.
The database was created with funding from NSF EIA-0196217.


\bibliographystyle{IEEEtran}
\bibliography{references}

\end{document}